\documentclass[a4paper]{jpconf}
\usepackage{graphicx}
\usepackage{amsmath}
\usepackage{mathrsfs}
\usepackage{subfig}
\usepackage{hyperref}

\begin{document}
\title{Topological defects with power-law tails}

\author{R V Radomskiy$^1$, E V Mrozovskaya$^1$, V A Gani$^{1,2}$, I C Christov$^3$}

\address{$^1$National Research Nuclear University MEPhI (Moscow Engineering Physics Institute), Kashirskoe highway 31, Moscow, 115409, Russia}
\address{$^2$National Research Centre Kurchatov Institute, Institute for Theoretical and Experimental Physics, Bolshaya Cheremushkinskaya str. 25, Moscow, 117218, Russia}
\address{$^3$School of Mechanical Engineering, Purdue University, West Lafayette, Indiana 47907, USA}

\ead{VAGani@mephi.ru}

\begin{abstract}
We study interactions of kinks and antikinks of the $(1+1)$-dimensional $\varphi^8$ model. In this model, there are kinks with mixed tail asymptotics: power-law behavior at one infinity versus  exponential decay towards the other. We show that if a kink and an antikink face each other in way such that their power-law tails determine the kink--antikink interaction, then the force of their interaction decays slowly, as some negative power of distance between them. We estimate the force numerically using the collective coordinate approximation, and analytically via Manton's method (making use of formulas derived for the kink and antikink tail asymptotics).
\end{abstract}

\section{Introduction}

Field-theoretic models with polynomial self-interaction are of growing interest in various areas of modern physics, from cosmology and high energy physics to condensed matter theory \cite{vilenkin01,manton,khare}.

In $(1+1)$-dimensional models of a real scalar field with a high-order polynomial self-interaction potential, there exist topological solutions of the type of kinks, which can possess tails such that either or both decay as power laws towards the asymptotic states connected by the kink \cite{khare,lohe}. Properties of kinks with exponential tail asymptotics (e.g., such as those arising as solutions to the sine-Gordon, $\varphi^4$ or $\varphi^6$ model) are well-understood \cite{GaKuPRE,aek01,christov01,GaKuLi,weigel02,GaLeLi,GaLeLiconf,dorey}. In particular, we know a lot about kink-antikink interactions in such models. Meanwhile, interactions of kinks with power-law tails have not been studied in such detail.

The study of some properties and interactions of plane domain walls in $(2+1)$ and $(3+1)$ dimensional worlds can often be reduced to studying the properties and interactions of kinks in $(1+1)$ dimensions \cite{GaLiRa,GaLeRaconf,GaKuYadFiz,Lensky,GaKsKuYadFiz01,GaKsKuYadFiz02}.

In this paper, we show that power-law asymptotics lead to long-range interaction between kinks and antikinks --- specifically, the force of the interaction can decay slowly, as some negative power of kink-antikink separation. This is a crucial difference from the (``classical'') case of kinks with exponential tail asymptotics. In the latter case, the kink-antikink interaction force always decays exponentially. 

\section{A $(1+1)$-dimensional $\varphi^8$ model featuring kinks with power-law tails}

\begin{figure}[h]
\centering
\includegraphics[scale=0.3]{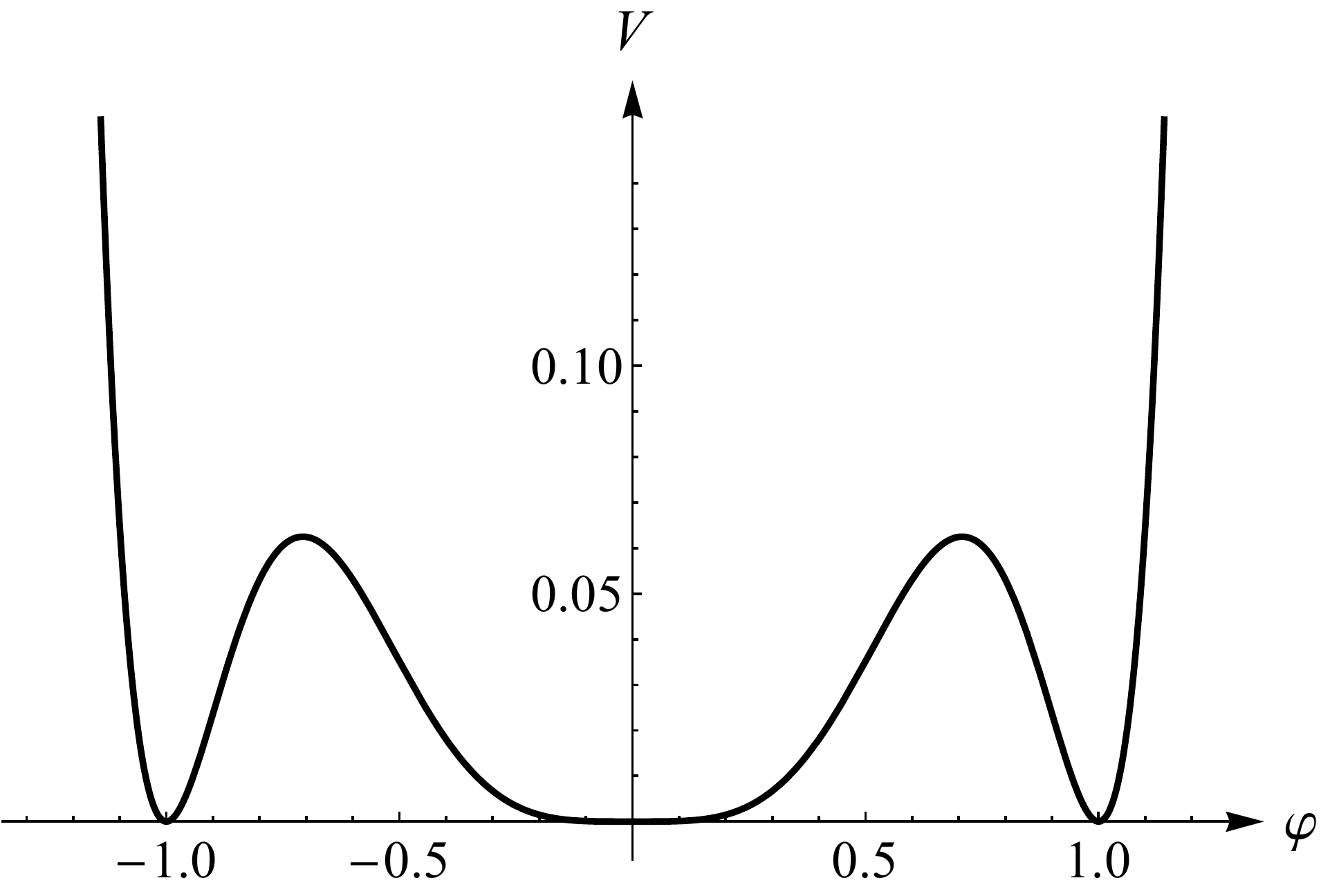}
\caption{Potential \eqref{eq:potential8} of the chosen $\varphi^8$ model.}
\label{fig:potential}
\end{figure}

Consider the $(1+1)$-dimensional $\varphi^8$ model \cite{khare,lohe}, given by the Lagrangian field density
\begin{equation}\label{eq:largang}
\mathscr{L}=\frac{1}{2} \left( \frac{\partial\varphi}{\partial t} \right)^2-\frac{1}{2} \left( \frac{\partial\varphi}{\partial x} \right) ^2-V(\varphi),
\end{equation}
where the potential of self-interaction is
\begin{equation}\label{eq:potential8}
V(\varphi)=\varphi^4(1-\varphi^2)^2,
\end{equation}
as depicted in figure \ref{fig:potential}. The potential \eqref{eq:potential8} has three degenerate minima: $\tilde{\varphi}_1=-1$, $\tilde{\varphi}_2=0$, and $\tilde{\varphi}_3=1$, such that  $V(\tilde{\varphi}_1)=V'(\tilde{\varphi}_1)=V(\tilde{\varphi}_2)=V'(\tilde{\varphi}_2)=V(\tilde{\varphi}_3)=V'(\tilde{\varphi}_3)=0$.

Due to the Lorentz invariance of $(1+1)$-dimensional field theories generated by \eqref{eq:largang} \cite{manton}, we restrict to static solution without loss of generality. Static solutions, which are called kinks \cite{manton}, interpolate between neighboring degenerate minima of the potential (vacua states of the model). We use the following notation for kinks: $\varphi_{(-1,0)}(x)$ denotes the kink connecting the vacua $\tilde{\varphi}_1=-1$ and $\tilde{\varphi}_2=0$. We can also say that this kink belongs to the ``topological sector'' $(-1,0)$, and similarly for the other kinks of the model.

\begin{figure}[h]
  \centering
  \subfloat[kink $\varphi_{(0,1)}(x)$]{\includegraphics[width=0.4\textwidth]{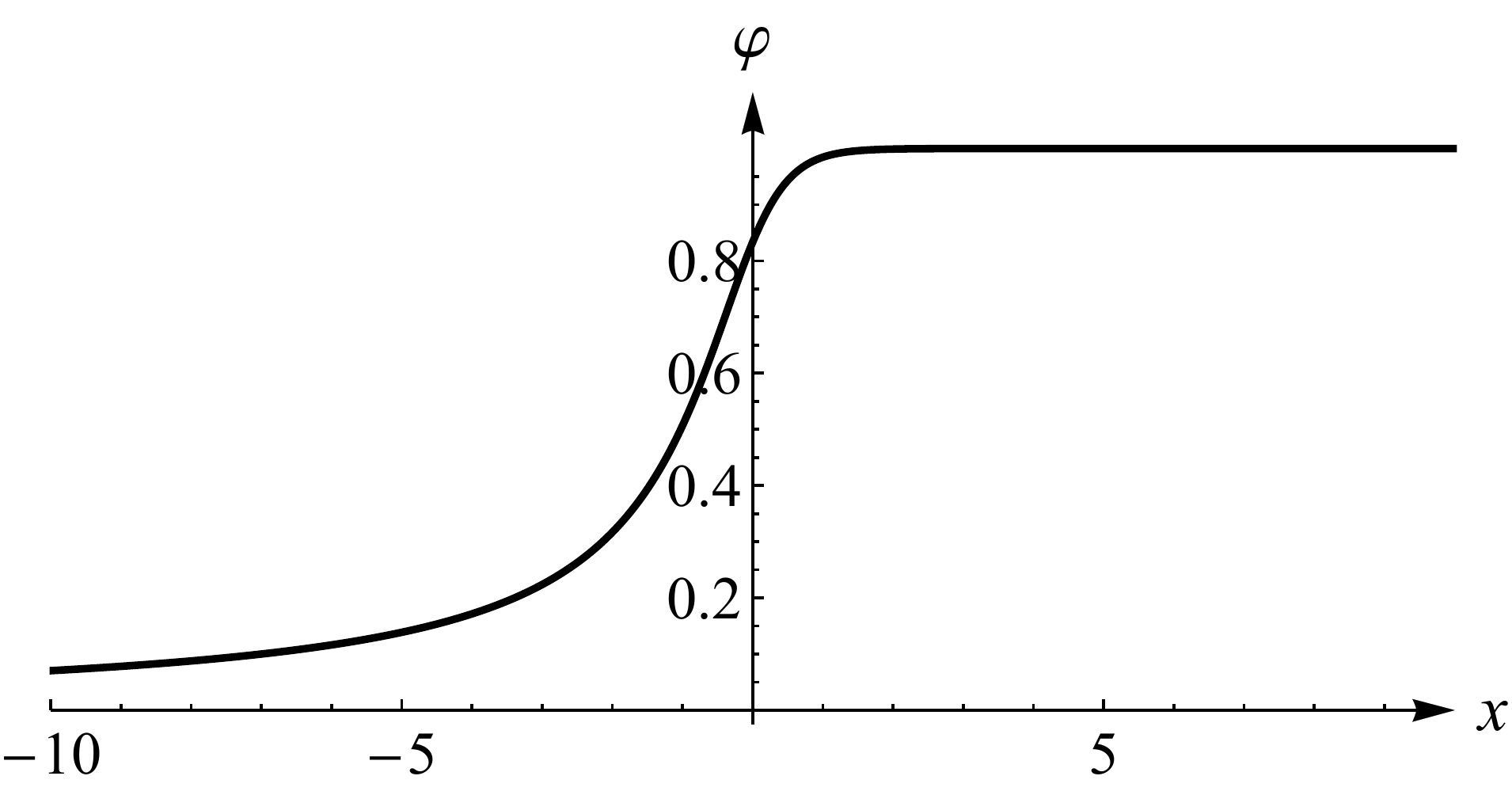}\label{fig:kinkU}}
  \hspace{15mm}
  \subfloat[kink $\varphi_{(-1,0)}(x)$]{\includegraphics[width=0.4\textwidth]{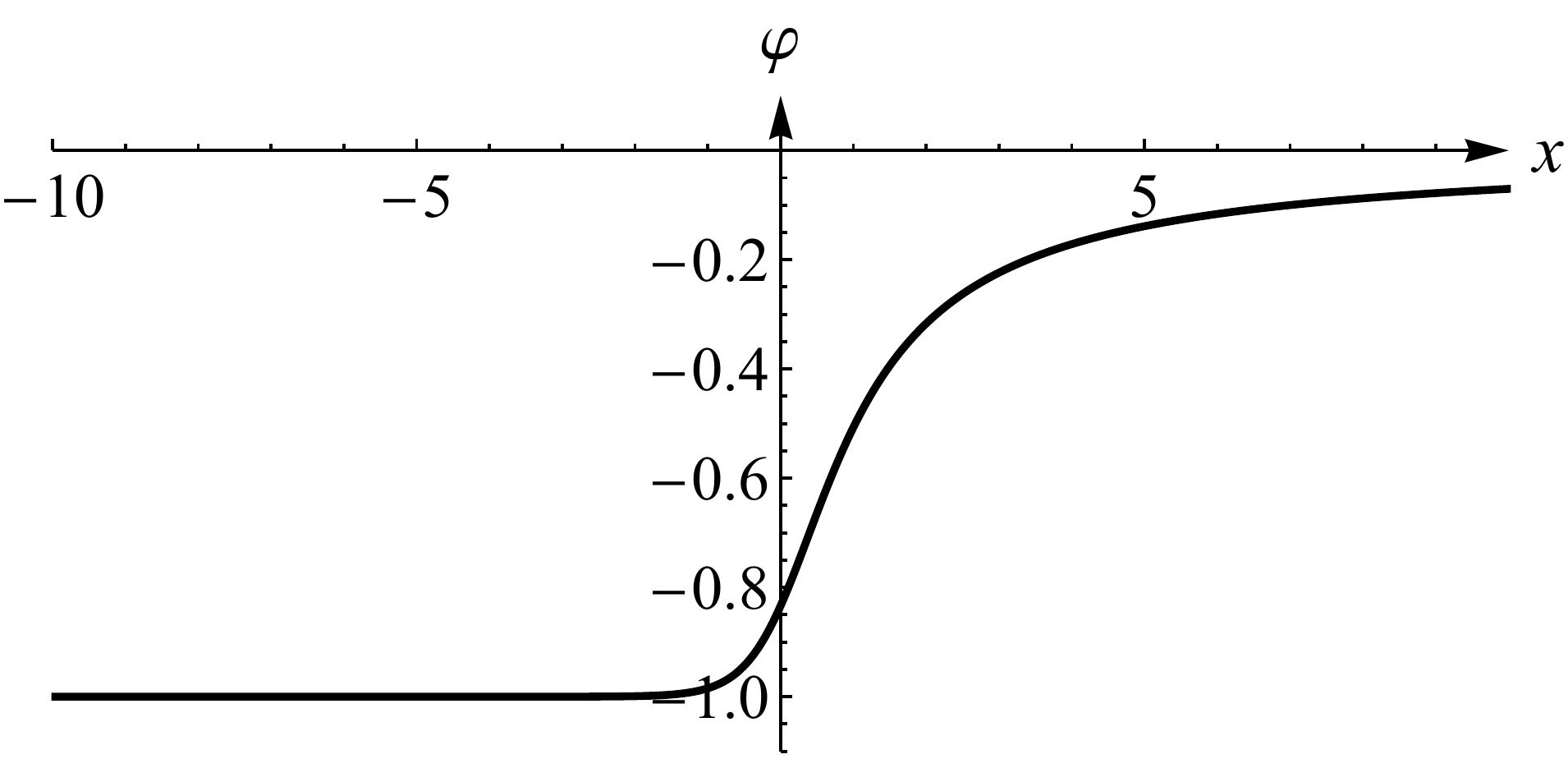}\label{fig:kinkD}}
  \caption{The two kink solutions \eqref{eq:kinks} of the chosen $\varphi^8$ model with potential \eqref{eq:potential8}.}
  \label{fig:kinks}
\end{figure}

The $\varphi^8$ model with the potential \eqref{eq:potential8} has two kink solutions, $\varphi_{(-1,0)}(x)$ and $\varphi_{(0,1)}(x)$ shown in figures \ref{fig:kinkU} and \ref{fig:kinkD}, respectively, both of which exhibit mixed power-law and exponential tail asymptotics. To each kink there is also a corresponding antikink. The kink $\varphi_{(-1,0)}(x)$ has power-law asymptotics at $x\to +\infty$, while the kink $\varphi_{(0,1)}(x)$ has power-law asymptotics at $x\to -\infty$. All kink solutions of the chosen model can easily be obtained in the implicit closed form \cite{khare,GaLeLi}:
\begin{equation}\label{eq:kinks}
2\sqrt2x=-\frac{2}{\varphi}+\ln\frac{1+\varphi}{1-\varphi}.
\end{equation}
Figure \ref{fig:kinks} illustrates the two distinct kink solutions.
(The corresponding expressions for antikinks can be obtained from \eqref{eq:kinks} via the transformation $x\mapsto -x$.)
Standard Taylor series expansions reveal that the tail asymptotics of the kinks given by equation \eqref{eq:kinks} are:
\begin{alignat}{2}
\varphi_{(-1,0)}(x) &\sim-1+\frac{2}{e^2}\: e^{2\sqrt{2}\: x}, &\qquad x\to -\infty.\label{eq:kink1_asymp_minus}\\
\varphi_{(-1,0)}(x) &\sim-\frac{1}{\sqrt{2}\: x}, &\qquad x\to +\infty.\label{eq:kink1_asymp_plus}\\
\varphi_{(0,1)}(x) &\sim-\frac{1}{\sqrt{2}\: x}, &\qquad x\to -\infty.\label{eq:kink2_asymp_minus}\\
\varphi_{(0,1)}(x) &\sim 1-\frac{2}{e^2}\: e^{-2\sqrt{2}\: x}, &\qquad x\to +\infty.\label{eq:kink2_asymp_plus}
\end{alignat}
Clearly, the kinks $\varphi_{(-1,0)}(x)$ and $\varphi_{(0,1)}(x)$ approach the vacuum state $\tilde{\varphi}_2=0$ very slowly (as $1/x$) at $x\to+\infty$ and $x\to-\infty$, respectively. Meanwhile, the approach to the vacua $\tilde{\varphi}_1=-1$ and $\tilde{\varphi}_3=1$ is exponential.

\section{Long-range interaction between a  kink and an antikink}
Consider a static configuration of a kink centered at some point $x=-\xi$ and an antikink centered at $x=+\xi$. Then, $\xi$ is the half-distance between the kink and the antikink. Our goal is to find how the force of kink-antikink interaction, in the chosen $\varphi^8$ model, depends upon $\xi$. Here,  by the force of interaction we mean the force produced on the kink by the antikink. To estimate this force, we use two methods: the collective coordinate approximation (see, e.g., \cite{manton,aek01,GaKuLi,weigel02} and the references therein for details) and Manton's method (see, e.g., \cite{manton,kks04}  and the references therein for details).

\subsection{Collective coordinate approximation}

\begin{figure}[h]
  \centering
  \subfloat[configuration \eqref{eq:ansatz1} for $\xi=15$]{\includegraphics[width=0.4\textwidth]{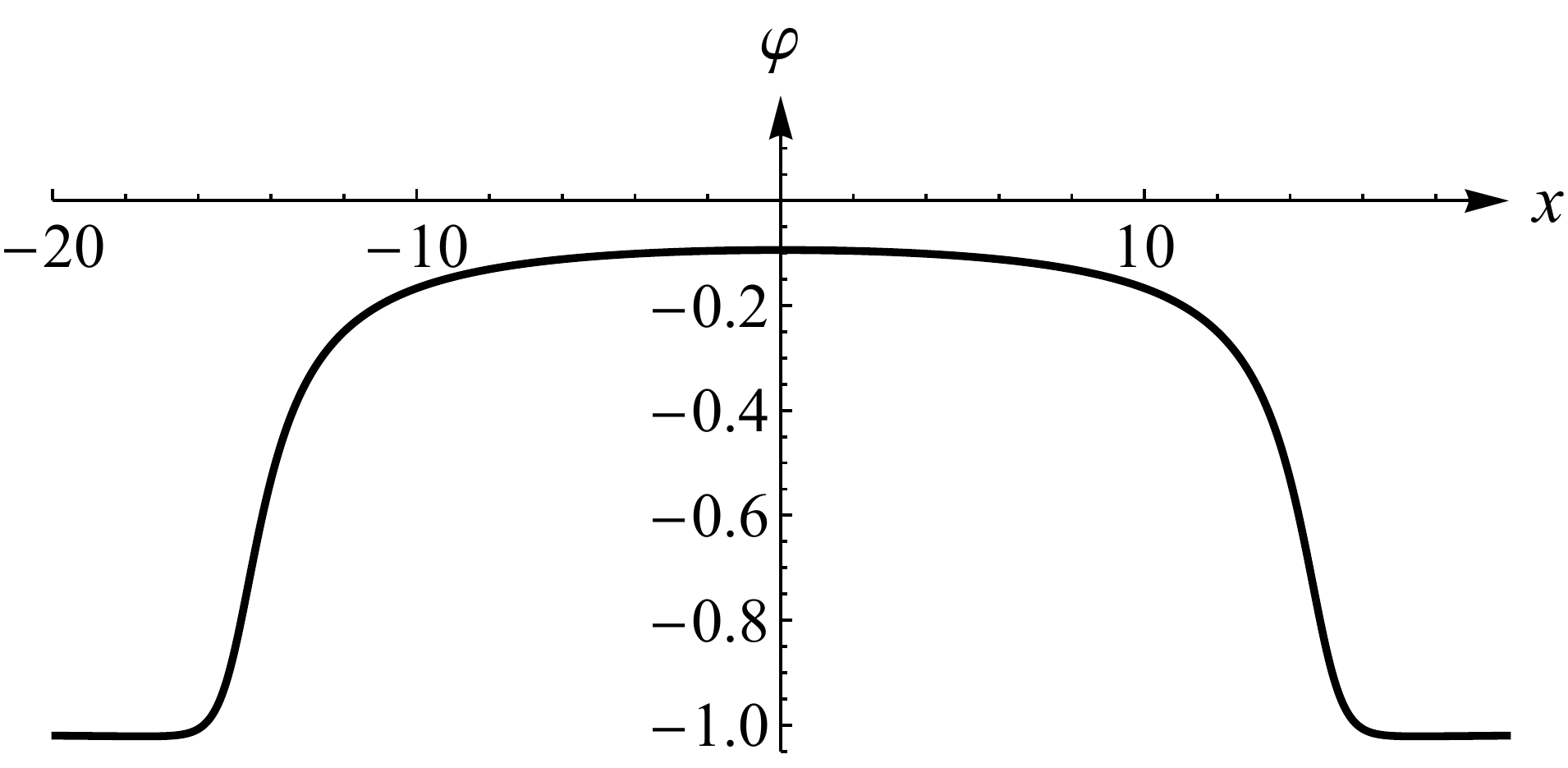}\label{fig:confD}}
  \hspace{15mm}
  \subfloat[configuration \eqref{eq:ansatz2} for $\xi=10$]{\includegraphics[width=0.4\textwidth]{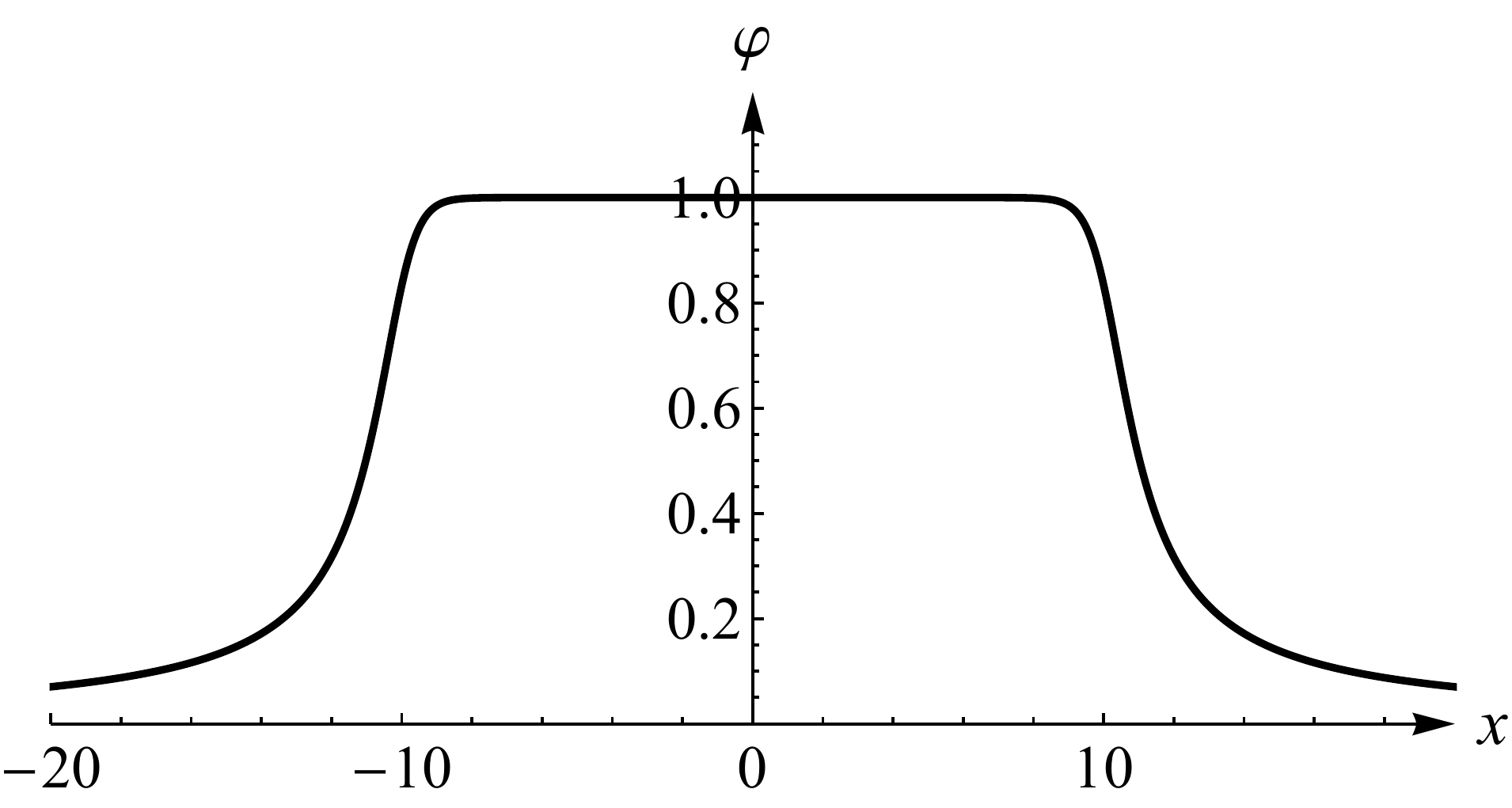}\label{fig:confU}}
  \caption{Kink-antikink configurations given by (a) equation \eqref{eq:ansatz1} and (b) equation \eqref{eq:ansatz2}.}
\end{figure}

In order to find the interaction force in the case of power-law tails, we assume the following ansatz for the kink-antikink field configuration:
\begin{equation}\label{eq:ansatz1}
\varphi(x;\xi)=\varphi_{(-1,0)}(x+\xi)+\varphi_{(0,-1)}(x-\xi),
\end{equation}
as shown in figure \ref{fig:confD}. By a standard calculation \cite{aek01}, we find that the effective potential $U_{\scriptsize\mbox{eff}}(\xi)$ and effective force $F(\xi)$ of the interaction are, respectively,
\begin{equation}\label{eq:U_eff}
U_{\scriptsize\mbox{eff}}(\xi)=\int_{-\infty}^{+\infty}\left[\frac{1}{2}\left( \frac{\partial\varphi}{\partial x} \right)^2 + V(\varphi)\right]dx, \qquad F(\xi)=\frac{dU_{\scriptsize\mbox{eff}}}{d\xi}.
\end{equation}
Here, $F(\xi)$ is the projection of the force onto the $x$-axis. Notice that we do not write the minus sign in front of the derivative of the effective potential because we are calculating the force on the left kink. Thus, a positive value of $-{dU_{\scriptsize\mbox{eff}}}/{d\xi}$ means that the force directed to the left, i.e.\ having a negative projection onto the $x$-axis. This corresponds to repulsion between the kink and antikink. Therefore, the second formula in \eqref{eq:U_eff} is written such that $F(\xi)>0$ corresponds to attraction, and $F(\xi)<0$ corresponds to repulsion.

In figure \ref{fig:forcepow} we show the dependence $F$ upon $\xi$ for the field configuration \eqref{eq:ansatz1}. It is seen that the kink and antikink repel each other, and the force falls off slowly with increasing separation.

Let us now consider the following kink-antikink field configuration:
\begin{equation}\label{eq:ansatz2}
\varphi(x,\xi)=\varphi_{(0,1)}(x+\xi)+\varphi_{(1,0)}(x-\xi)-1,
\end{equation}
which is shown in figure \ref{fig:confU}. Again, we seek to determine $F(\xi)$. In this case, the kink and antikink are turned to each other by the exponential tails. The numerically calculated force of interaction is presented in figure \ref{fig:forceexp}. From this figure, it is seen that the kink and antikink attract, and the force falls off quickly with increasing separation.

\subsection{Manton's method}

Within the framework of Manton's method \cite{manton}, the force on the kink is given by the time derivative of the momentum on the semi-infinite interval $-\infty<x\leq 0$. At large $\xi$ this method allows us to use the tail asymptotics for the kink and antikink, which were given in \eqref{eq:kink1_asymp_minus}--\eqref{eq:kink2_asymp_plus}, to approximate the integrands in the various integrals with respect to $x$.

Specifically, using the tail asymptotics  \eqref{eq:kink1_asymp_plus} of the kink and the tail asymptotics of the corresponding antikink ($x\mapsto-x$), we estimate the force of repulsion for the configuration \eqref{eq:ansatz1} (i.e., the kink and antikink are turned to each other by the power-law tails) to be 
\begin{equation}\label{eq:force_power}
F_\mathrm{M}(\xi) \sim \frac{4}{\xi^4},\qquad \xi \gg 1.
\end{equation}
In the case that the kink and antikink are turned to each other by their exponential tails, i.e.\ the configuration \eqref{eq:ansatz2}, the force of attraction is estimated to be
\begin{equation}\label{eq:force_exp}
F_\mathrm{M}(\xi) \sim \frac{64}{e^4} e^{-4\sqrt{2}\:\xi},\qquad \xi \gg 1.
\end{equation}

\begin{figure}[h]
  \centering
  \subfloat[force of interaction for configuration \eqref{eq:ansatz1}]{\includegraphics[scale=0.4]{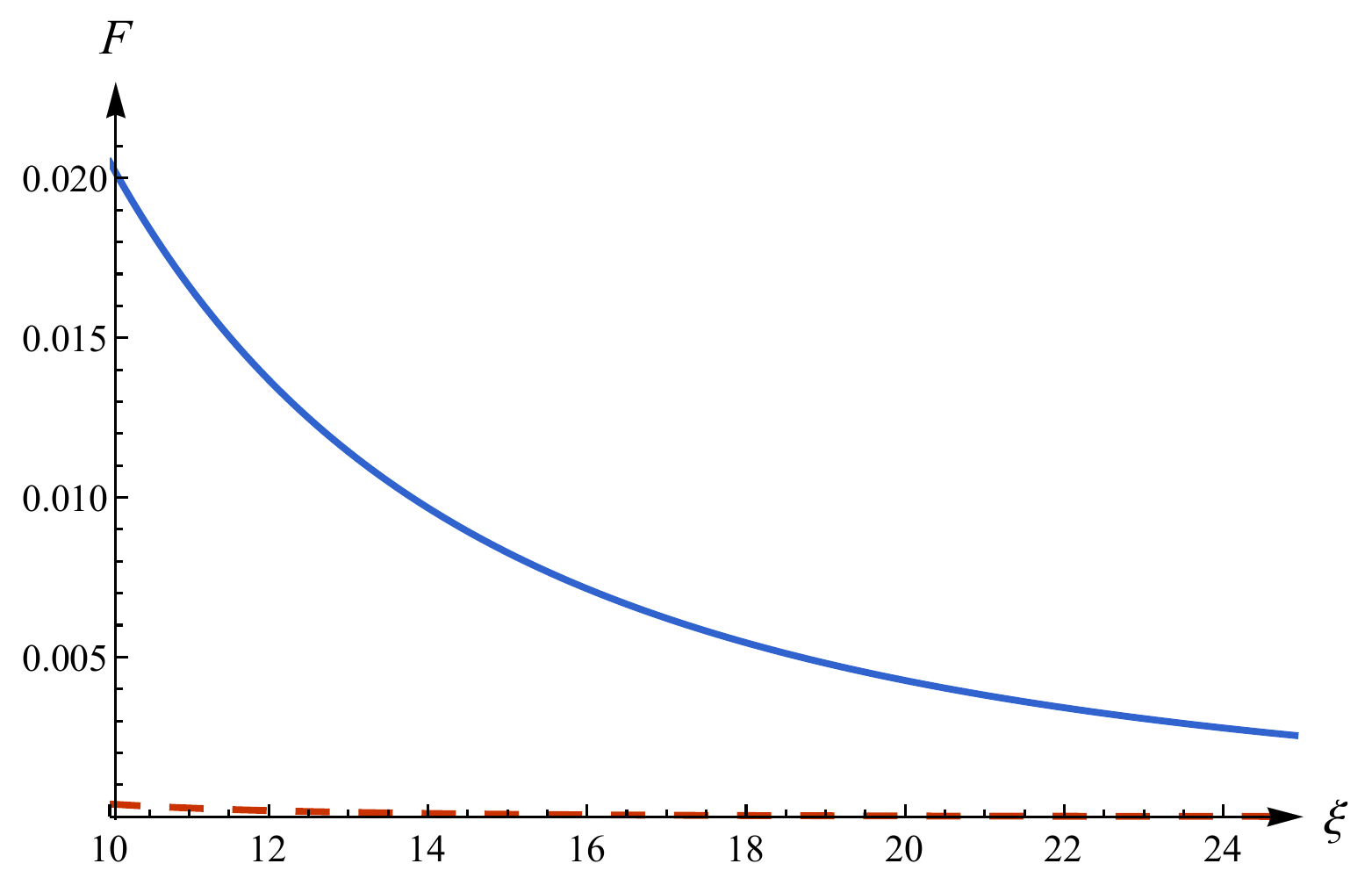}
\label{fig:forcepow}}
  \hspace{15mm}
  \subfloat[force of interaction for configuration \eqref{eq:ansatz2}]{\includegraphics[scale=0.4]{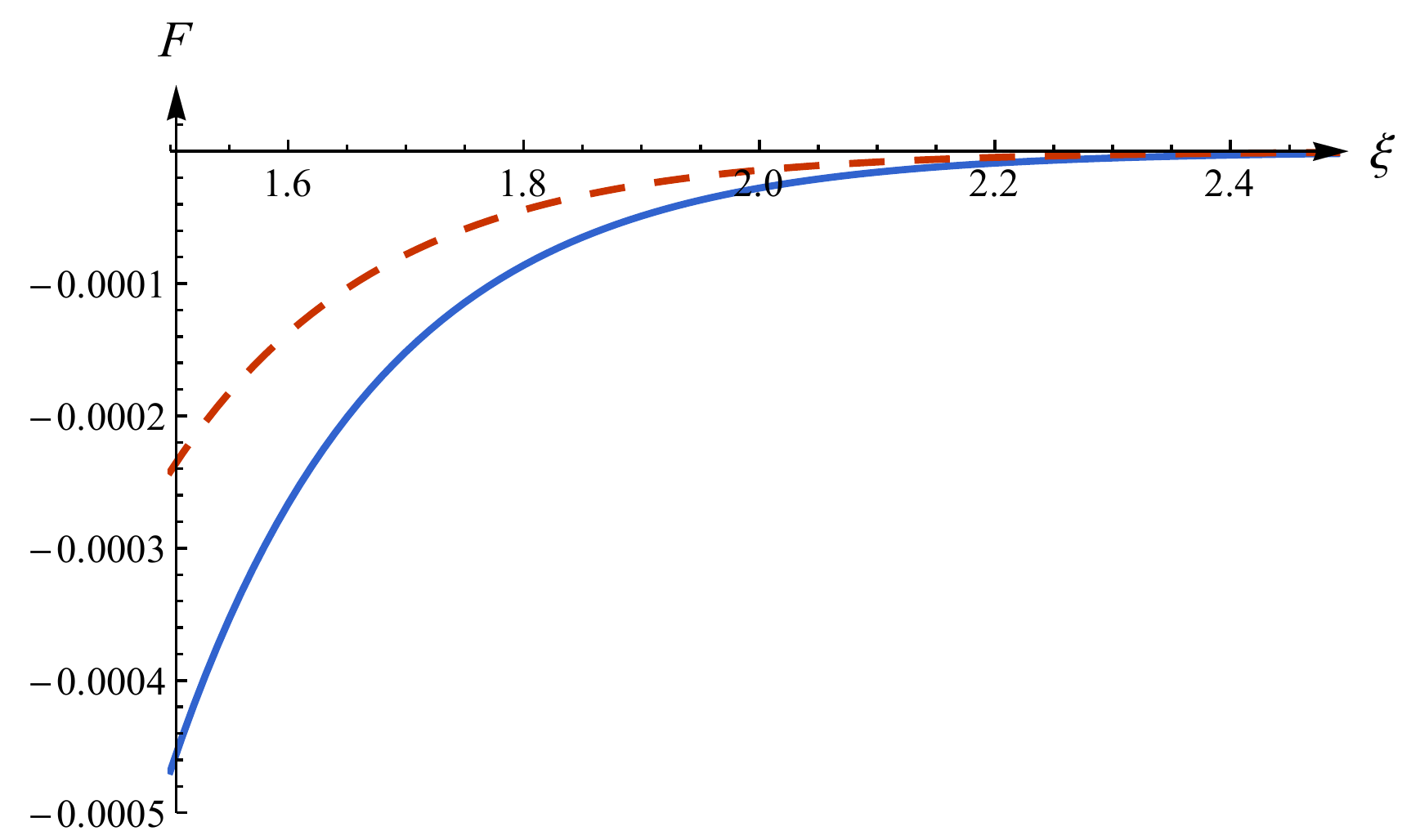}\label{fig:forceexp}}
  \caption{The force of the interaction $F(\xi)$ between a kink and an antikink, as a function of their half-separation $\xi$, for (a) the configuration \eqref{eq:ansatz1} (power-law tails) and (b) the configuration \eqref{eq:ansatz2} (exponential tails). The solid curves correspond to $F$ estimated by the collective coordinate approach (i.e., equation \eqref{eq:U_eff}), while the dashed curves correspond to $F$ estimated by Manton's method (i.e., equations \eqref{eq:force_power} and \eqref{eq:force_exp}).}
  \label{fig:forces}
\end{figure}

The predictions of \eqref{eq:force_power} and \eqref{eq:force_exp} (Manton's method) are compared to the corresponding result from \eqref{eq:U_eff}, as shown in figure \ref{fig:forces}. While in the case of exponential tail asymptotics (figure \ref{fig:forceexp}) we observe good agreement, the case of power-law tails (figure \ref{fig:forcepow}) shows considerable disagreement between the two curves. Nevertheless, the force of interaction for the case of power-law tails calculated from the collective coordinate approximation does appear to decay as $\xi$ to a negative integer power close to $-4$ (as predicted by \eqref{eq:force_power}), however the prefactor is off by orders of magnitude. We intend to investigate the origin of this discrepancy, and whether it is a fundamental limitation of Manton's method, in future work.

\section{Conclusion}

Within a specific $\varphi^8$ model, we have shown that if a kink and an antikink interact via tails that decay as power-laws, then a long-range interaction appears in the system --- the force of the kink-antikink interaction decays much more slowly than in the (``usual'') case of  exponentially decaying tails. 

Using Manton's method \cite{manton}, we have calculated the asymptotic dependence of the force of interaction on the half-distance $\xi$ between the kink and antikink. In the case of power-law tails, the force decays as $\xi^{-4}$ for $\xi\gg 1$. Meanwhile, for the case of exponential tails, this force decays off exponentially as $ e^{-4\sqrt{2}\xi}$ for $\xi\gg 1$.

Furthermore, we calculated the force between the kink and antikink numerically using the collective coordinate approximation. The results obtained by both methods are in a good agreement in the case of exponential tails. The origin of the discrepancy in the case of power-law tails will be investigated in future work.

Finally, note that understanding the long-range interactions of topological defects with power-law tails is important because such long-range interactions between kinks and antikinks can have key consequences for the dynamics of domain walls and other similar structures in field-theoretical models with polynomial self-interaction.

\section*{Acknowledgments}
This work was performed within the framework of the Center of Fundamental Research and Particle Physics supported by MEPhI Academic Excellence Project (contract No.~02.03.21.0005, 27.08.2013).


\section*{References}
\begin{thebibliography}{20}

\bibitem{vilenkin01}
Vilenkin A and Shellard E P S 2000, {\it Cosmic Strings and Other Topological Defects} (Cambridge: Cambridge University Press)

\bibitem{manton}
Manton N and Sutcliffe P 2004 {\it Topological Solitons} (Cambridge: Cambridge University Press)

\bibitem{khare}
Khare A, Christov I C and Saxena A 2014 \href{http://dx.doi.org/10.1103/PhysRevE.90.023208}{{\it Phys.~Rev.} E {\bf 90} 023208} (\href{http://arxiv.org/abs/1402.6766}{{\it arXiv}:1402.6766 [math-ph]})

\bibitem{lohe}
Lohe M A 1979 \href{http://dx.doi.org/10.1103/PhysRevD.20.3120}{{\it Phys.~Rev.} D {\bf 20} 3120}

\bibitem{GaKuPRE}
Gani V A and Kudryavtsev A E 1999 \href{http://dx.doi.org/10.1103/PhysRevE.60.3305}{{\it Phys.~Rev.} E {\bf 60} 3305} (\href{http://arxiv.org/abs/cond-mat/9809015}{{\it arXiv}:cond-mat/9809015})

\bibitem{aek01}
Belova T I and Kudryavtsev A E 1997 \href{http://dx.doi.org/10.1070/PU1997v040n04ABEH000227}{{\it Phys.~Usp.} {\bf 40} 359}

\bibitem{christov01}
Christov I and Christov C I 2008 \href{http://dx.doi.org/10.1016/j.physleta.2007.08.038}{{\it Phys.~Lett.~A} {\bf 372} 841} (\href{http://arxiv.org/abs/nlin/0612005}{{\it arXiv}:nlin/0612005 [nlin.PS]})

\bibitem{GaKuLi}
Gani V A, Kudryavtsev A E and Lizunova M A 2014 \href{http://dx.doi.org/10.1103/PhysRevD.89.125009}{{\it Phys.~Rev.} D {\bf 89} 125009} (\href{http://arxiv.org/abs/1402.5903}{{\it arXiv}:1402.5903 [hep-th]})

\bibitem{weigel02}
Takyi I and Weigel H 2016 \href{http://dx.doi.org/10.1103/PhysRevD.94.085008}{{\it Phys.~Rev.} D {\bf 94} 085008} (\href{http://arxiv.org/abs/1609.06833}{{\it arXiv}:1609.06833 [nlin.PS]})

\bibitem{GaLeLi}
Gani V A, Lensky V and Lizunova M A 2015 \href{http://dx.doi.org/10.1007/JHEP08(2015)147}{{\it JHEP} {\bf 08} 147} (\href{http://arxiv.org/abs/1506.02313}{{\it arXiv}:1506.02313 [hep-th]})

\bibitem{GaLeLiconf}
Gani V A {\it et al.} 2016 \href{http://dx.doi.org/10.1088/1742-6596/675/1/012019}{{\it J.~Phys.: Conf.~Ser.} {\bf 675} 012019} (\href{http://arxiv.org/abs/1602.02636}{{\it arXiv}:1602.02636 [hep-th]})

\bibitem{dorey}
Dorey P {\it et al.} 2011 \href{http://dx.doi.org/10.1103/PhysRevLett.107.091602}{{\it Phys.~Rev.~Lett.} {\bf 107} 091602} (\href{http://arxiv.org/abs/1101.5951}{{\it arXiv}:1101.5951 [hep-th]})

\bibitem{GaLiRa}
Gani V A, Lizunova M A and Radomskiy R V 2016 \href{http://dx.doi.org/10.1007/JHEP04(2016)043}{{\it JHEP} {\bf 04} 043} (\href{https://arxiv.org/abs/1601.07954}{{\it arXiv}:1601.07954 [hep-th]})

\bibitem{GaLeRaconf}
Gani V A, Lizunova M A and Radomskiy R V 2016 \href{http://dx.doi.org/10.1088/1742-6596/675/1/012020}{{\it J.~Phys.: Conf.~Ser.} {\bf 675} 012020} (\href{http://arxiv.org/abs/1602.04446}{{\it arXiv}:1602.04446 [hep-th]})

\bibitem{GaKuYadFiz}
Gani V A and Kudryavtsev A E 2001 \href{http://dx.doi.org/10.1134/1.1423755}{{\it Phys.~Atom.~Nucl.} {\bf 64} 2043} (\href{http://arxiv.org/abs/hep-th/9904209}{{\it arXiv}:hep-th/9904209}, \href{http://arxiv.org/abs/hep-th/9912211}{{\it arXiv}:hep-th/9912211})

\bibitem{Lensky}
Lensky V A, Gani V A and Kudryavtsev A E 2001 \href{http://dx.doi.org/10.1134/1.1420436}{{\it J.~Exp.~Theor.~Phys.} {\bf 93} 677} (\href{http://arxiv.org/abs/hep-th/0104266}{{\it arXiv}:hep-th/0104266})

\bibitem{GaKsKuYadFiz01}
Gani V A, Ksenzov V G and Kudryavtsev A E 2010 \href{http://dx.doi.org/10.1134/S1063778810110104}{{\it Phys.~Atom.~Nucl.} {\bf 73} 1889} (\href{http://arxiv.org/abs/1001.3305}{{\it arXiv}:1001.3305 [hep-th]})

\bibitem{GaKsKuYadFiz02}
Gani V A, Ksenzov V G and Kudryavtsev A E 2011 \href{http://dx.doi.org/10.1134/S1063778811050085}{{\it Phys.~Atom.~Nucl.} {\bf 74} 771} (\href{http://arxiv.org/abs/1009.4370}{{\it arXiv}:1009.4370 [hep-ph]})

\bibitem{kks04}
Kevrekidis P G, Khare A and Saxena A 2004 \href{http://dx.doi.org/	
10.1103/PhysRevE.70.057603}{{\it Phys.~Rev.} E {\bf 70} 057603} (\href{http://arxiv.org/abs/nlin/0410045}{{\it arXiv}:nlin/0410045 [nlin.PS]})

\end{thebibliography}
\end{document}